\documentclass[11pt,twoside]{article}

\usepackage{graphics,graphicx}
\usepackage{asp2006}
\usepackage{epsf}
\usepackage{psfig}
\usepackage{lscape}
\newcommand{\feiif}{ [Fe~II]} 
\newcommand{\mic}{$\mu$m}
\newcommand{\spitzer}{\textit{Spitzer}}

\begin{document}
\title{Dust Formation Observed in Young Supernova Remnants with Spitzer}   
\author{Rho, J. (SSC/Caltech), Reach, W. T. (Planck/Caltech), Tappe, A. (CFA/Harvard), 
Rudnick, L. (UMN), Kozasa, T. (Hakkaido U.),
Hwang, U. (GSFC/NASA), Andersen, M. (SSC/Caltech), Gomez, H. (Cardiff), DeLaney, T. (MIT), 
Dunne, L. (U. Nottingham), Slavin, J. (CFA/Harvard)}   

\begin{abstract} 
We present dust features and masses observed in young supernova
remnants (SNRs) with Spitzer IRS mapping and staring observations of
four youngest supernova remnants: SNR 1E102.2-7219 (E0102) in the SMC,
Cas A and G11.2-0.3 in our Galaxy, and N132D in the LMC. The spectral mapping data
revealed a number of dust features which include 21 micron-peak dust
and featureless dust in Cas A and 18-micron peak dust in E0102 and
N132D. The 18 micron-peak feature is fitted by a mix of MgSiO$_3$ and
solid Si dust grains, while the 21-micron peak dust is by a mix of
silicates and FeO; we also explore dust fitting using Continuous
Distribution of Ellipsoid grain models. We report detection of CO
fundamental band from Cas A in near-infrared. We review dust
features observed and identified in other SNRs.  The dust emission is
spatially correlated with the ejecta emission, showing dust is formed
in SN ejecta. The spectra of E0102 show rich gas lines from ejecta
including strong ejecta lines of Ne and O, including two [Ne III] lines
and two [Ne V] lines which allow us to diagnostic density and
temperature of the ejecta and measure the ejecta masses.  E0102 and
N132D show weak or lacking Ar, Si, and Fe ejecta, whereas the young
Galactic SNR Cas A show strong Ar, Si, and S and weak Fe. We 
discuss  compositions and masses of dust and association with those of
ejecta and finally, dust contribution from SNe to early Universe.

\end{abstract}



\subsection*{1. Introduction}

Meteoritic and astronomical stuides show that presolar, cosmic grains
condense in  the dense, warm stellar winds of evolved stars  and
in supernova explosions. Mantles of the pre-existing dust in molecular
clouds  are vaporized as the forming stars and planetary system heat
them. A small fraction of the dust survived solar system formation
without alteration, protected inside asteroids.  The most abundant
presolar grains are SiC, nanodiamonds, amorphous silicates, forsterite
and enstatite,  and corundum (Al$_2$O$_3$) \citep{messenger}. 
Some isotopic anomalies of heavy elements in 
meteorites have been attributed to the dust that had condensed deep
within expanding supernovae.

Recent deep sub-mm observations have also shown there to be galaxies and
QSOs with very large dust masses ($>$10$^8$ $\rm{M_{\odot}}$) at z$>5$
(Bertoldi et al. 2003). The timescales for low-mass stars to release
their dust are too long to explain these high redshift systems. In
contrast, supernovae produce copious amounts of heavy elements and
release them on short timescales (Morgan \& Edmunds 2003; Dwek et al.
2008). Theoretical modeling of the conditions in the supernova ejecta
indicate that Type-II SNe are sources of dust formation and should be
able to produce substantial quantities of dust, on the order of a solar
mass per explosion \citep{den03,todini01,noz03}.  Yet until very
recently, there existed little observational evidence that this actually
occurs.

There is now clear evidence for dust formation in core-collapse
supernovae, but the {\em quantity} of dust formed within SNe ejecta  is
still a subject of debate. For SN1987A, this includes dust emission,
dust absorption and a drop in line intensities for the refractory
elements that signals that dust is being formed (Kozasa et al. 1989). 
Detections of SiO and CO fundamentals with
\spitzer\ observations futher provide evidence of dust formation in SNe 
\citep{kot06}. For the young Galactic SNR, Cas\,A, ISO observations gave
evidence for the association of the dust with the ejecta by requiring a
mixture of dust grains that are not typical of the ISM. Submillimeter
observations of Cas\,A and Kepler with SCUBA \citep{dun03, mor03a}
suggest the presence of large amounts of cold dust ($\sim
0.3-2\,\rm{M_{\odot}}$ at 15--20 K), but with some controversy related
to foreground material.  \cite{kra04} showed that much of the 160$\mu$m
emission observed with {\it Spitzer}  is foreground material, suggesting
there is no cold dust in Cas A. \cite{wil05}, however, used CO emission
towards the remnant to show that up to about a solar mass of dust could
still be associated with the ejecta rather than with the foreground
material.

We present {\it Spitzer} IRS spectral mapping of the entire SNRs of Cas A,
E0102, and N132D and an IRS staring mode data of G11.2-0.3, 
and near-infrared observations of Cas A for the first overtone CO detection.
We identify the dust composition associated  with the ejecta and discuss dust
formation in SNe with an estimate of the total mass of freshly
formed dust.

\subsection*{2. Cas A} 

Cas A is one of the youngest Galactic SNRs, with an age of 335 yr
attributed to a SN explosion in AD 1671. The progenitor of Cas A is
believed to be a Wolf-Rayet star with a mass of 15-30  M$_{\odot}$
\citep{you06} 
 We present a short summary of {\it Spitzer} Infrared
Spectrograph (IRS) mapping  observations of Cas A which were published
in Rho et al. (2008), and we added spectral fitting with
the dust models of  continuous distributions of ellipsoidals (CDE).

\begin{figure}[!h]
\hbox{
\psfig{figure=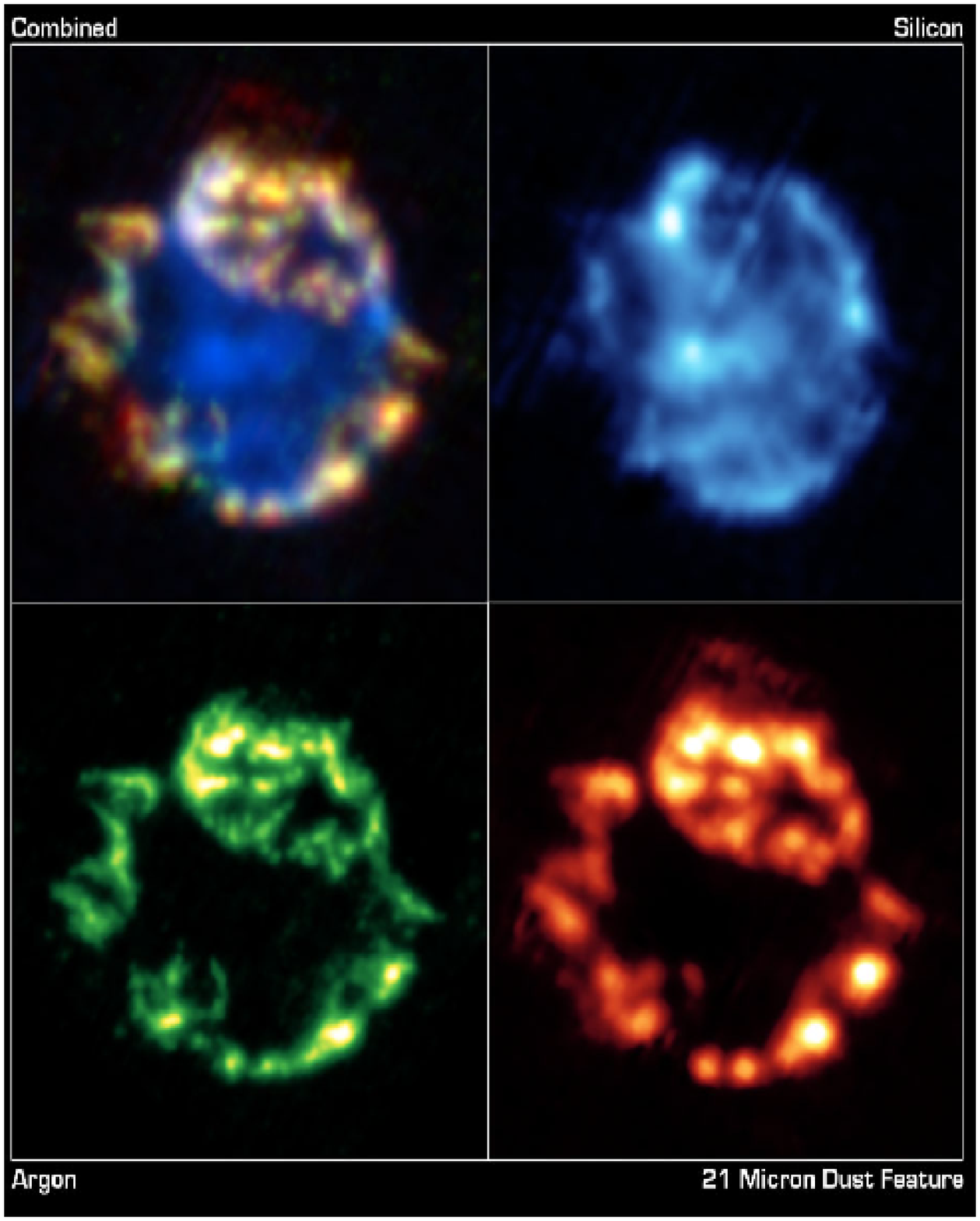,angle=0,height=7truecm,width=6.1truecm}
\vbox{
\psfig{figure=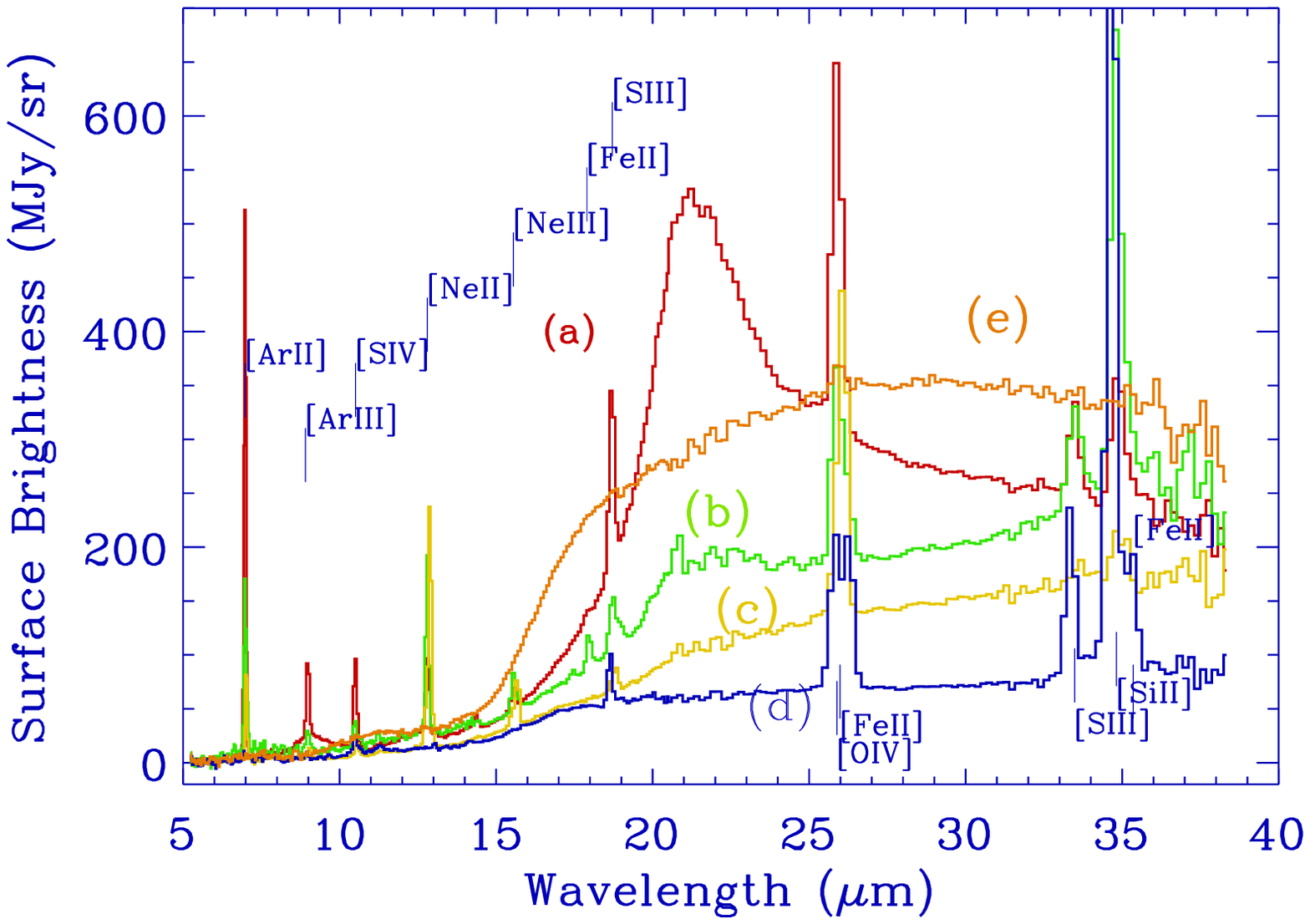,height=3.5truecm,width=5truecm}
\psfig{figure=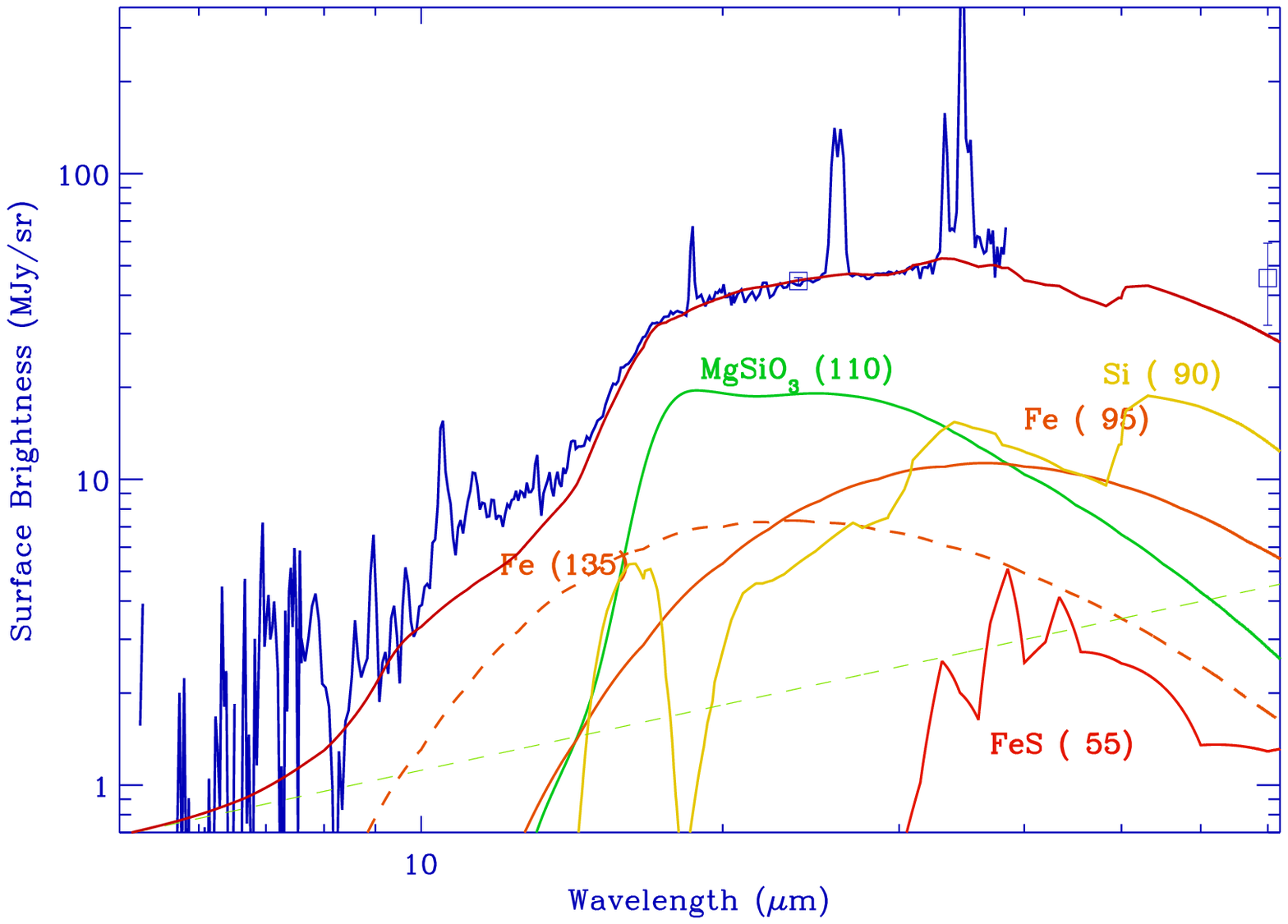,height=3.5truecm,width=5truecm}
}
}
\caption{(a: left) The 21 $\mu$m dust map of Cas A shows remarkable
similarity to the  Ar ejecta map, showing that dust is formed in the
ejecta. (b: top right) Representative set of {\it Spitzer} IRS spectra
of Cas A with rich ejecta lines,  
The dust feature at 21 $\mu$m often accompanied
by a silicate emission feature at 9.8 $\mu$m with strong Ar lines (red,
curve a) and weak-21 $\mu$m dust with relatively strong Ne lines
compared with Ar (green curve, b). Featureless spectra include the
continuous rising spectra (c and d). (c: bottom right) Featureless
dust Spectrum: the continuum can be fit with MgSiO$_3$ and Fe
accompanied with  S, Si, and O/Fe lines.
\label{casafig1} }
\end{figure}

\begin{figure}[!h]
\hbox{
\psfig{figure=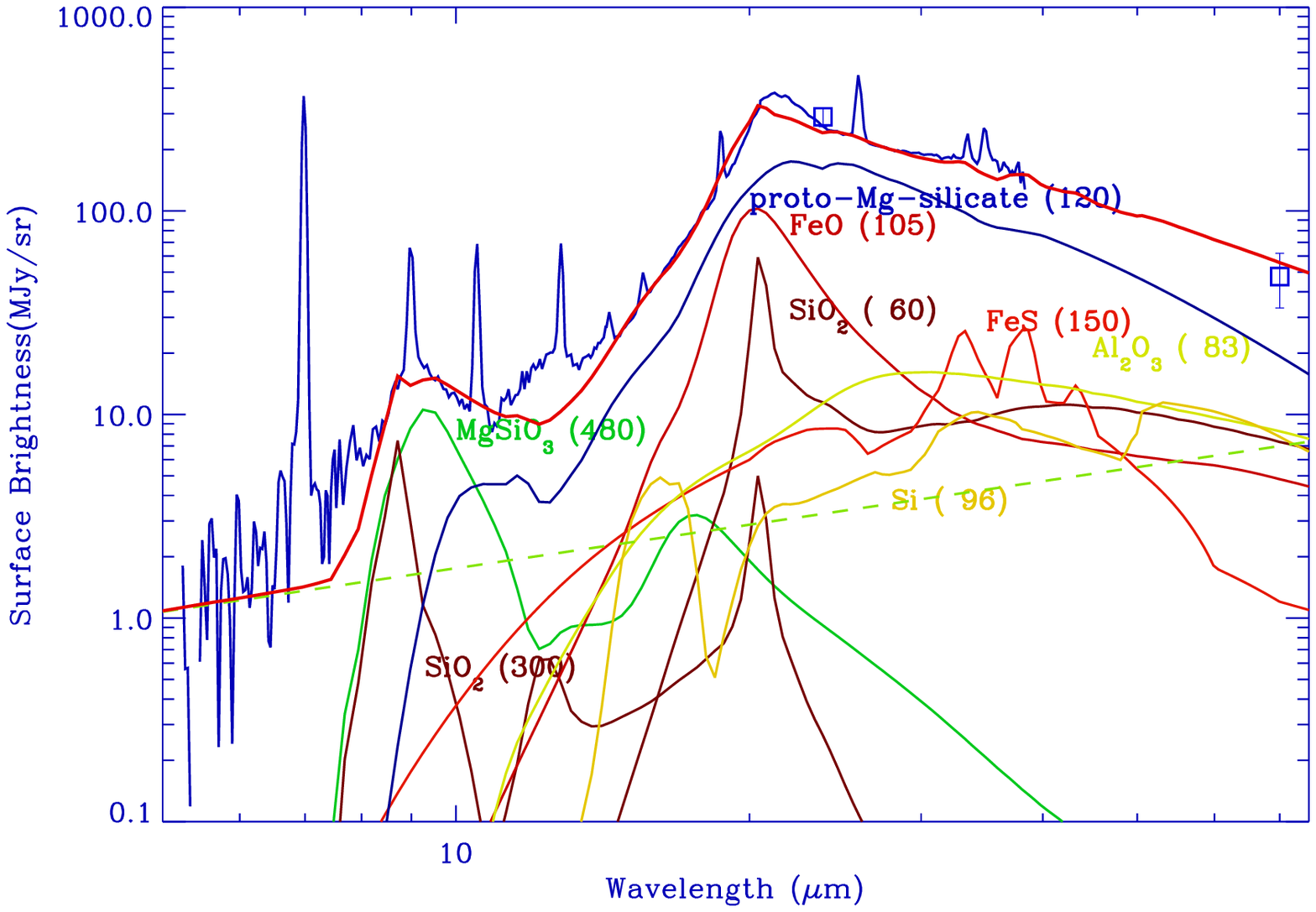,angle=0,height=4.5truecm}
\psfig{figure=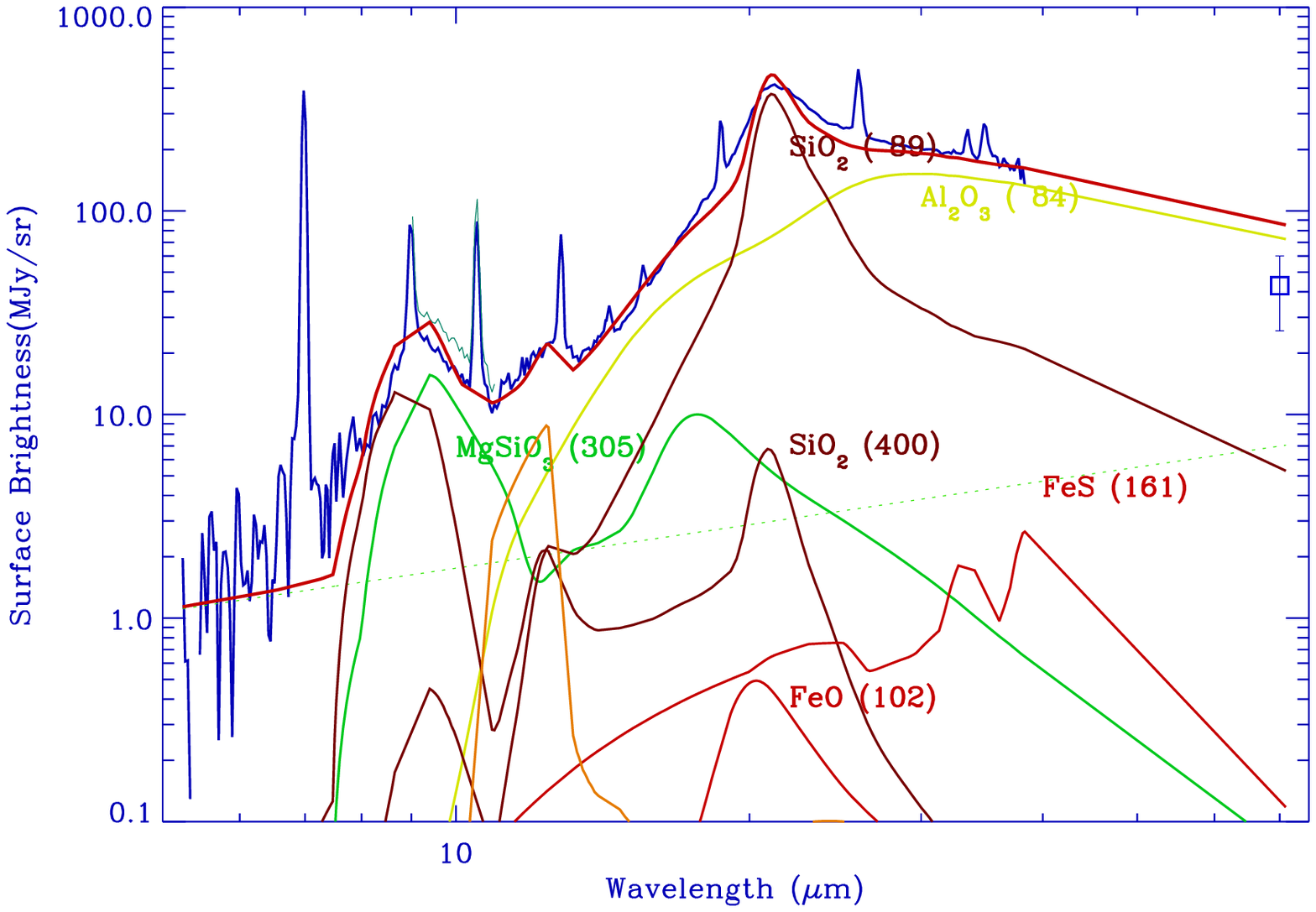,angle=0,height=4.5truecm}
}
\caption{(a: left) 21 $\mu$m-peak dust spectrum  superposed on the dust fit  of
Model A: A {\it Spitzer} IRS spectrum towards a bright part of the
northern shell fitted with dust compositions of Mg proto-silicate,
MgSiO$_3$,  SiO$_2$, FeO,  and Al$_2$O$_3$. The compositions suggest
that the dust forms around inner-oxygen and S-Si layers.
(b: right) Spectral fitting with continuous distribution of ellipsoidal models of
SiO$_2$, MgSiO$_3$ as well as other composition of Al$_2$O$_3$, FeS and FeO, showing
CDE model of SiO$_2$ can reasonably well reproduce the 21$\mu$m-peak dust without
the  Mg proto-silicate. }
\label{casafig2}
\end{figure}

Gas lines of Ar, Ne, O, Si, S and Fe, and dust continua were strong for
most  positions. We identify three distinct ejecta dust  populations
based on their continuum shapes.  The dominant dust continuum  shape
exhibits a strong peak at 21 $\mu$m.  A line-free map of 21 $\mu$m-peak 
dust made from the 19-23 $\mu$m range closely  resembles the [Ar~II],
[O~IV], and [Ne~II] ejecta-line maps as shown in Fig. \ref{casafig1}a,
implying that dust is freshly formed in the ejecta.   We identify three
distinct classes of dust associated  with the ejecta as shown in Fig.
\ref{casafig1}b. Spectral fitting of the 21$\mu$m-peak dust  implies
the  presence of SiO$_2$, Mg protosilicates, and FeO grains in these
regions in Fig. \ref{casafig2}a.    The silicate composition is
responsible for the 21 $\mu$m peak, suggesting that the dust forms
around the inner-oxygen and S-Si layers and is consistent with Ar being
one of the oxygen burning products.  We also include amorphous
MgSiO$_3$  (480 K) and SiO$_2$ (300 K) to account for the emission 
feature around the 9.8 $\mu$m. The composition of  the low temperature
(40-90 K) dust component necessary for  reproducing 70 $\mu$m is rather
unclear.  Either Al$_2$O$_3$ (80 K) (Model A in Table 1 of  Rho et al.
2008) or  Fe (100 K) (Model B in Table 1) can fit equally well.

The  second dust type exhibits a rising continuum up to 21 $\mu$m and
then flattens  thereafter.  This ``weak 21 $\mu$m'' dust is likely
composed of Al$_2$O$_3$ and C grains.  The third dust continuum shape is
featureless with a gently  rising spectrum and is likely composed of
MgSiO$_3$ and either Al$_2$O$_3$ or Fe grains (Fig. \ref{casafig1}c).
Using the least massive composition  for each of the three dust classes
yields a total mass of 0.02 M$_\odot$. Using the most-massive
composition yields a total mass of 0.054 M$_\odot$.  The primary
uncertainty in the total  dust mass stems from the selection of the dust
composition  necessary for fitting  the featureless dust as well as 70
$\mu$m flux.    The freshly formed dust mass derived from Cas A may be
able to explain the lower limit on the dust masses in high redshift
galaxies when assuming there is no significant dust destruction.

We also performed spectral fitting of the 21$\mu$m-peak dust with CDE models
for SiO$_2$ and MgSiO$_3$ and the spherical model for the rest of dust composition. 
The 21$\mu$m peak dust is reasonably well fit by with SiO$_2$  and Al$_2$O$_3$ 
in Fig. \ref{casafig2}b without the composition of
Mg proto-silicate and with much less contribution of FeO.
In addition, previously unidentified 11-12.5$\mu$m feature can be
fit with the CDE model of SiC, which is one of the important pre-solar grains.
Slight excess of model at 10$\mu$m absorption feature can be compensated
when we use the extinction of \cite{chiar07} in Fig. \ref{casafig2}b. 
The estimate of dust mass using the CDE fittting of SiO$_2$ slightly increases the mass,
but within the range of the previous estimate of Rho et al. (2008).

\subsection*{3. SNR 1E0102.2-72.3 (E0102)} 

We present \spitzer\ IRS and IRAC observations of the young supernova
remnant E0102 (SNR 1E0102.2-7219) in the Small Magellanic Cloud.   We
performed an IRS staring observation toward the southeastern shell  of
E0102 (R.A.\ $01^{\rm h} 04^{\rm m} 04.04^{\rm s}$ and Dec.\ 
$-72^\circ$02$^{\prime} 00.5^{\prime \prime}$, J2000)
as a part of our Young SNR \spitzer\ GO program 
(PI: Rho).  The Long Low (LL: 15-40 $\mu$m) IRS data were taken on 
2005 August 14 with 6 cycles of 30 sec exposure time; this yields a 
total exposure time of 360 sec for the first and second staring 
positions.  The Short Low (SL: 5-15 $\mu$m) IRS observations were made 
with 3 cycles of 60 sec exposure time; this yields a total exposure 
time of 360 sec per sky position.  We also used archival IRS mapping
data from the IRS Legacy SMC program, which data were 6
times shallower than our GO IRS staring mode data, but still helpful to
generate line maps.

\begin{figure}[!h]
\hbox{
\psfig{figure=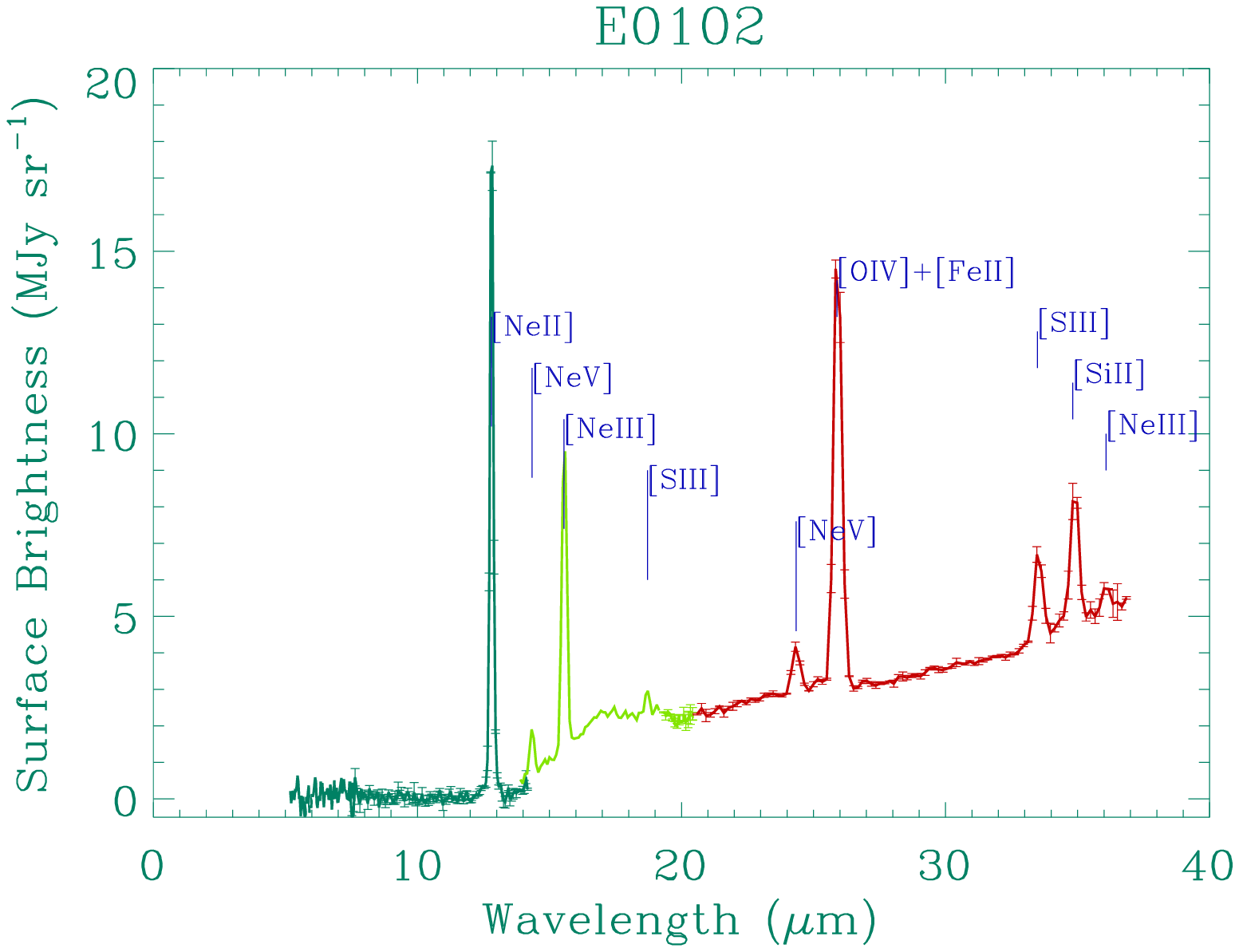,angle=0,height=5.truecm}
\psfig{figure=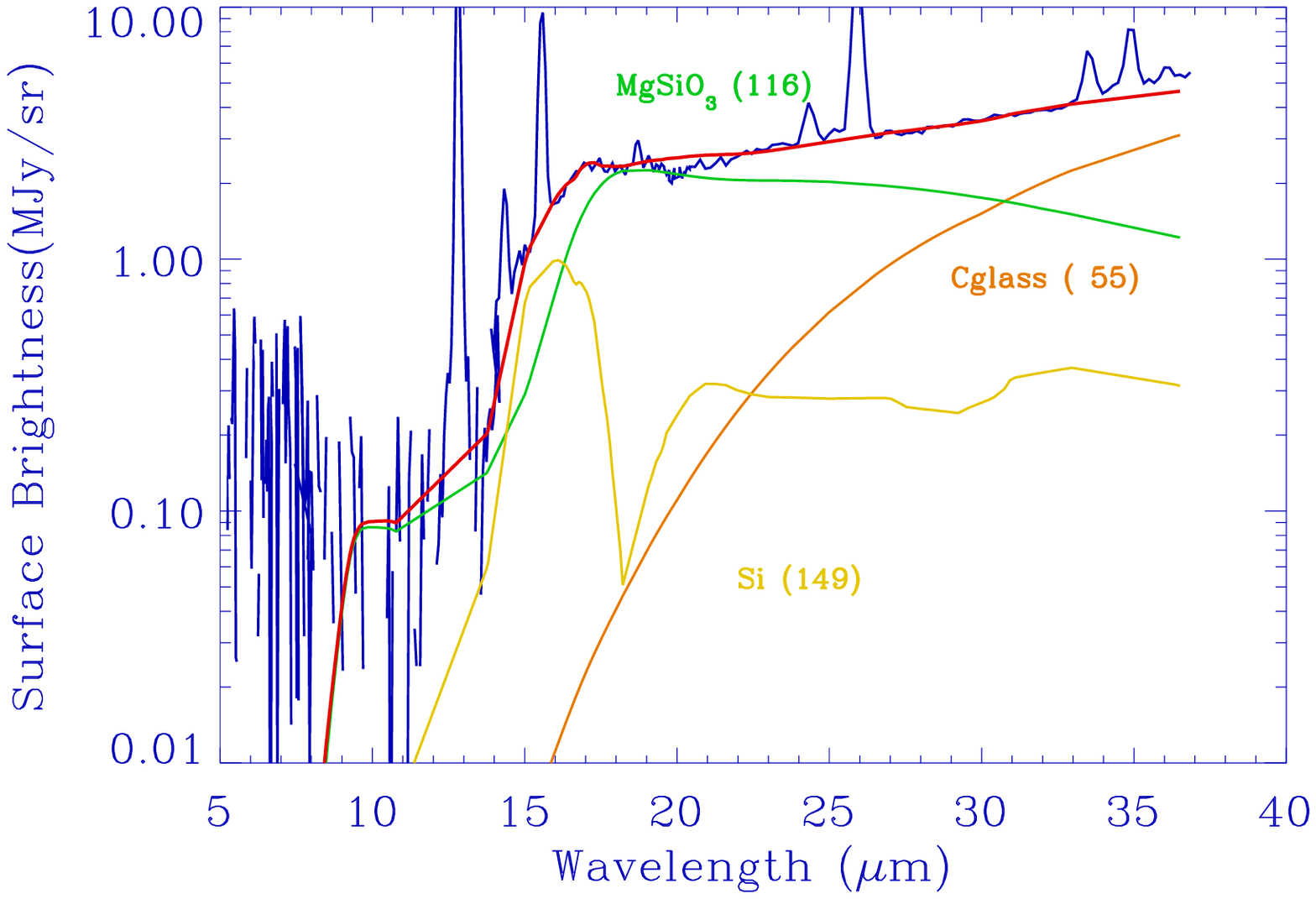,angle=0,height=5.truecm}
}
\caption{(a) {\it Spitzer} IRS spectrum of E0102.
(b) E0102 IRS dust spectrum superposed on a dust fit (Model A).  
 The continuum was fitted with dust compositions 
of  Si, MgSiO$_3$ and Carbon.  The compositions suggest that the dust 
forms around carbon-burning layers. The data and the total fit are shown 
in blue and thick red lines, respectively. Contributions from each type 
of dust are shown with the dust temperatures given in parentheses. The gas lines are  
also shown in the plot, but they are excluded in spectral fitting 
}
\label{E0102irs}
\end{figure}

The infrared spectra in Figure \ref{E0102irs} show  both line and
continuum. The detected lines include  strong ejecta lines of Ne and O,
including two [Ne III] lines at 15.5 and 36.0 \mic, and two [Ne V] lines
at 14.3 and  24.3 $\mu$m and weak Si and S lines. Unlike the young
Galactic SNR Cas A, E0102 lacks emission from Ar and Fe, both in the
infrared and at other wavelengths.  The  [Ne II]\ line at 12.8 \mic\ is
broad, with a velocity dispersion of  2,000-4,500 km s$^{-1}$, showing
that it originates in fast-moving  ejecta.

\noindent {\bf 3.1. Dust Spectral Fitting and Dust Mass of E0102}

\begin{figure}[!h]
\hbox{
\psfig{figure=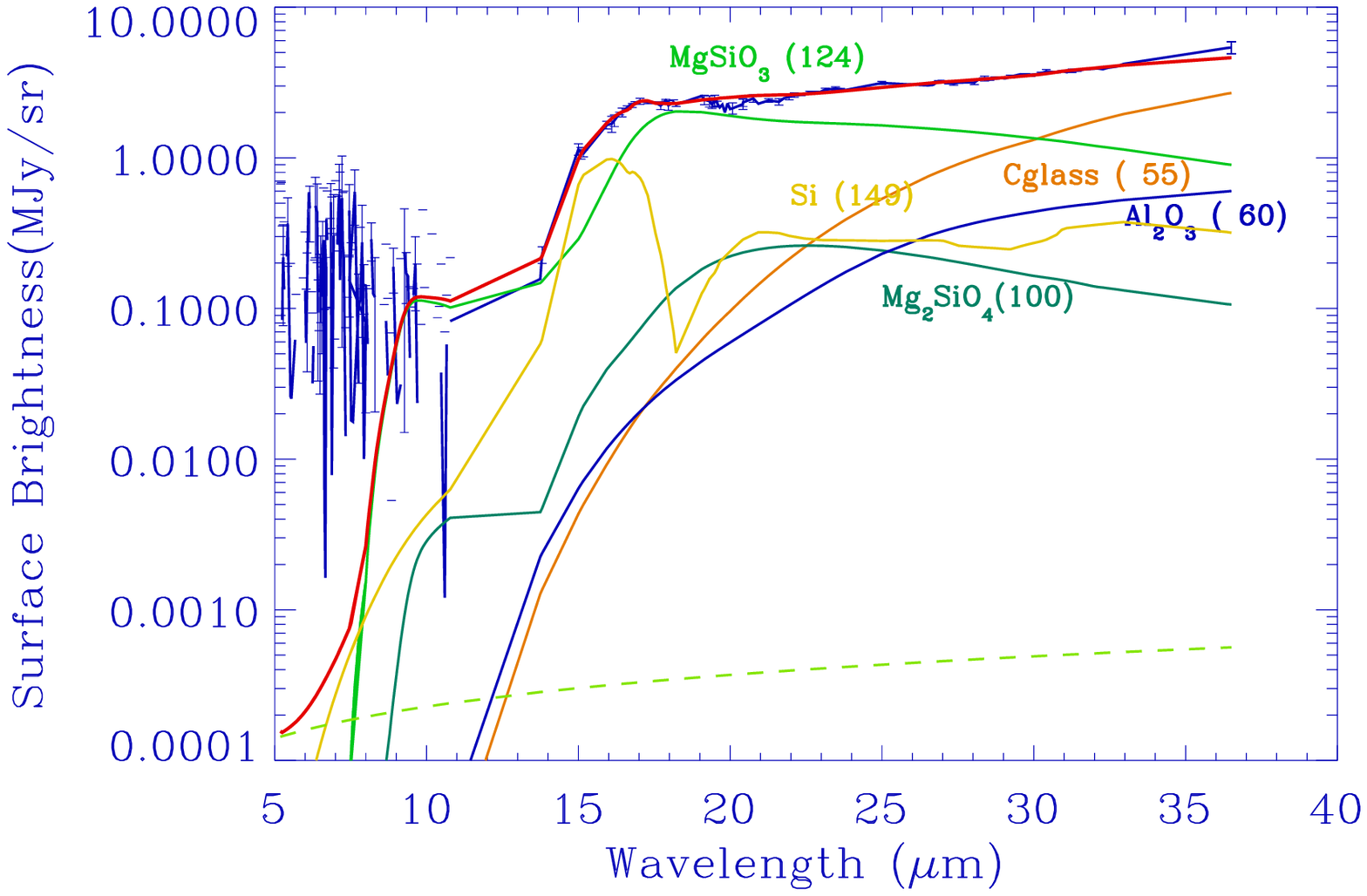,angle=0,height=5.truecm}
\psfig{figure=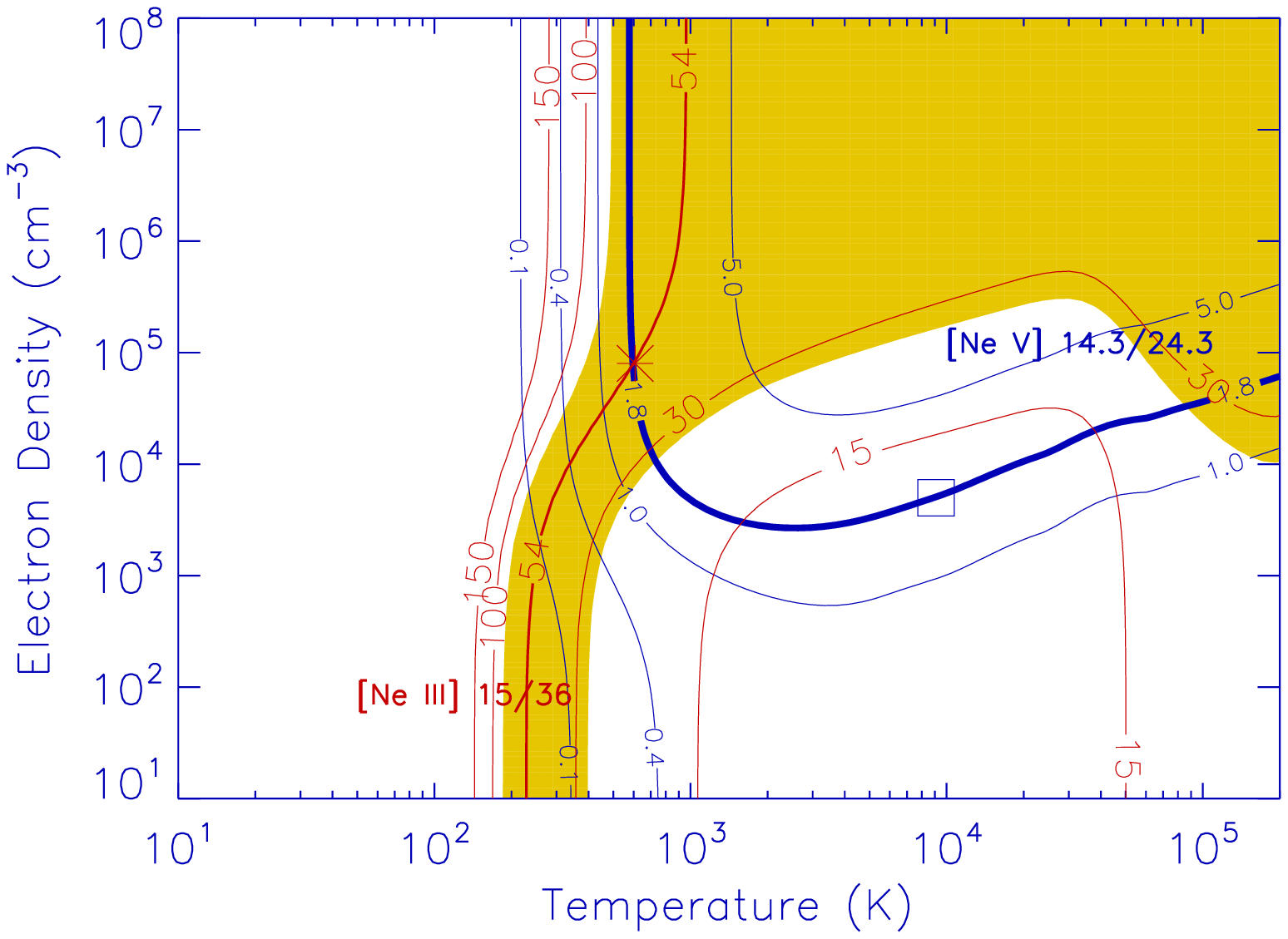,angle=0,height=5.truecm}
}
\caption{(a: left) The dust continuum superposed on a combination of dust models 
of Si, MgSiO$_3$, Al$_2$O$_3$, Mg$_2$SiO$_4$, and carbon.
(b: right)
Line diagnostic contours of the ratios of [Ne~III] 15/35\mic\  (red)
and [Ne~V] 14.3/24.3\mic\ (blue).  The observed ratios are  marked with
thick solid lines for the best value of the ratios  [Ne III]  = 54 and
[Ne V]\ = 1.76. Note that the density is an electron density.   The
larger shaded region shows the range  of temperatures and densities
allowed by errors for the [Ne II]  ratio, whereas   the allowed
physcial conditions of [Ne V] lines are marked with thick lines.
} 
\label{E0102dustfitcarbon}
\end{figure}

The IRS spectra of E0102.2-7219 have a prominent dust feature peaking
at 18 \mic\ which coincides spatially with the emission from the
infrared-emitting ejecta. This is clear evidence that dust is forming
in the ejecta of the supernova remnant E0102.  To determine the dust
composition and mass, we performed spectral fitting to the IRS dust
continuum using the deep IRS spectra shown in Figure \ref{E0102irs}. 
The spectrum shows a 18\mic-peak in the continuum which we attribute to
dust.  First, we estimated the contribution of the synchrotron emission
to the infrared spectrum using the radio fluxes (0.65 Jy at 408 MHz)
and spectral index (-0.70)  \citep{amy93}. The expected synchrotron
fluxes are (1.4 - 5.5)$\times 10^{-4}$ MJy sr$^{-1}$ between 5-40 \mic,
which is a very small contribution to the infrared continuum ($<$1\%,
see Fig.\ref{E0102dustfitcarbon}).

We performed spectral fitting to the IRS continua with the Planck function
$B_\nu(T)$ multiplied by the absorption efficiency ($Q_{abs}$) for
various dust compositions and the detailed
dust fitting technique and mass
estimation method is described in \cite{rho08}.
The dust compositions, temperatures, and masses obtained from the best
fits are summarized in Table \ref{tbestfit}.

\vskip -0.5truecm
\begin{table}[!h]
{\scriptsize
\caption{Properties of Freshly Formed Dust in E0102}
\label{tbestfit}
\begin{center}
\begin{tabular}{llllll}
\hline
Model ($\Delta \chi^2$)  & Dust Compositions (temperature [K], mass [M$_\odot$]) & Mass (M$_\odot$)\\
\hline
A (2.41)  &{\bf MgSiO$_3$ (116, 3.80E-5), Si (150, 6.20E-5)}, {\it C (55, 5.00E-3)} & 0.015 ($^a$5.12$\times 10^{-3}$)  \\ 
B (1.98)   &{\bf MgSiO$_3$ (117, 3.21E-5), Si (114, 4.4E-4)}, {\it Al$_2$O$_3$ ($\equiv$55,  2.06E-3)} & 0.007 ($^a$2.53$\times 10^{-3}$)\\
C (2.16)  & {\bf MgSiO$_3$ (124, 2.34E-5), Si (149, 6.3E-5)}, & \\
 & Al$_2$O$_3$ (60, 3.33eE-4), C (55, 4.350E-3), Mg$_2$SiO$_4$ (100, 1.10E-5) & 0.014 
($^a$4.80$\times 10^{-3}$) \\
\hline
\label{tbestfit}
\end{tabular}
\end{center}
}
\end{table}

{\vskip -0.7truecm}
The 18\mic-peak dust feature is fitted by a mix of MgSiO$_3$ and Si dust
grains, while the rest of the continuum requires either carbon (Model A;
Figure ~\ref{E0102dustfitcarbon}) or Al$_2$O$_3$ grains (Model B).  The
temperature of Al$_2$O$_3$ cannot be constrained by our spectra, which
lack long-wavelength data,  so we set the same temperature as the carbon
in Model A. We favor  carbon or Al$_2$O$_3$ over solid Fe dust, because
we expect carbon or  Al$_2$O$_3$ dust to be present where the Ne and O
ejecta lines are  dominant: Ne, Mg, and Al are all carbon-burning
nucleosynthesis  products \citep{woosley95}.  MgSiO$_3$ and Al$_2$O$_3$
are condensed in the  ejecta of Ne, Mg, and Al, which are also
carbon-burning products; the O  and Al are found in the outer layers of
ejecta, which is where Si and  carbon dust grains condense. The dust and
ejecta compositions seen in  E0102 are similar to those of the weak
21$\mu$m dust layer in Cas A  \citep{rho08}.  To explore the effect of
including many minerals, in  Model C (Figure \ref{E0102dustfitcarbon})
we performed fit including  MgSiO$_3$, Si, Al$_2$O$_3$, carbon  and
Mg$_2$SiO$_4$ (the most commonly expected dust compositions from Ne and
O ejecta; see  e.g.T01, N03).  Note that presolar Al$_2$O$_3$ grains
from meteorites have been inferred to be among most abundant
isotopically-enriched materials ejected by Type II SNe
\citep{clayton04}.  The mass is relatively well constrained by the 
observations (within 20-40\%).

The estimated dust masses within the IRS observed slit are
5.0$\times$10$^{-3}$ and 5.5$\times$10$^{-3}$ M$_{\odot}$ for Models A
and D, respectively.  To estimate the total dust mass, we correct a
factor of 2.9, the observed  (within the high-sensitivity spectral slit)
mass for the fraction of the SNR covered by the IRS slit using a
continuum map. The total dust masses for the entire SNR are then 0.015,
and 0.014 M$_{\odot}$ for Models A, and D, respectively. Our estimation
exceeds previous estimates (8$\times$10$^{-4}$ M$_{\odot}$; Stanimirovic
et al. 2005) by more than an order of magnitude because we have
accounted for dust compositions.

\noindent{\bf 3.2. Ne Line Diagnostics and Shock Models}

The ions Ne~III and Ne~V ions offer pairs of lines in the IR that are suitable
diagnostics for the density and temperature in emitting regions.  The measured
line flux ratios are [Ne~V] $\lambda$ 14.3/24.3$\mu$m = $1.76\pm$0.11, and 
[Ne~III] $\lambda$ 15.6/36$\mu$m $= 54.3\pm$27.4. To constrain temperatures
and densities, we calculate the line intensities and ratios of [Ne~V] and 
[Ne~III].  We solve the level population equations for collisional excitation
as a matrix using 5 energy levels for Ne~V and 3 levels for Ne~III.  The input
atomic data were taken from \cite{gri00}.  The line diagnostics are shown in
Figure \ref{E0102dustfitcarbon}. The electron density and temperature jointly
obtained from the ratios of [Ne~V]\ and  [Ne~III]\ are 7$\times$10$^4$
cm$^{-3}$ and $\sim$610 K, when assuming that both Ne~III and Ne~V gas come
from the same gas.  Such a high density seems to be characteristic of  SN
ejecta as observed in optical lines; \cite{chevalier78}, for example, suggested
that the optical [S~II]\ line in Cas~A originated from ejecta  with a density
of 10$^5$ cm$^{-3}$.  \cite{blair00}, however, find considerably lower densities
when matching optical lines observed in bright optically emitting knots in
E0102.

We run radiative shock model with a velocity range of 50 - 500 km s$^{-3}$,
and use basic parameters quoted in \cite{blair00} and \cite{hughes98} including
the abundances.  To calculate the hot, non-radiative region of the shock, we
calculate the steady shock profile, integrating the energy equation and
including the magnetic field, radiative cooling and non-equilibrium
ionization.  The cooled photoionized region of the shock requires a code that
can calculate the thermal balance at the low temperatures and high densities
of that gas and for this purpose we use Cloudy \citep{ferland06}.  We use the
EUV/X-ray flux generated in the hot post-shock region as the input ionizing
flux and use the results from Cloudy on the Ne ionization to explore the
parameters needed to match the  observed [Ne~II], [Ne~III], and [Ne~V] lines.
The shock model indicates that the [Ne~II] and [Ne~III] lines are mainly from
photoinization zone with a  cold temperature of 400 - 1000 K, while the [Ne~V]
comes from hotter density region. Following \cite{blair00} we calculate a
series of shocks with the same ram pressure but shock speeds ranging from 50 -
500 km s$^{-1}$ such that $n_0 v_\mathrm{shock}^2 = (16$ cm$^{-3})(100$ km
s$^{-1})^2$. Also following \cite{blair00} we assume a magnetic field strength
of $B_0 = 4 n_0(\mathrm{cm}^{-3})^{1/2}\; \mu$G.  The derived [Ne~II] column
densities of (2-40)$\times$10$^{14}$ cm$^{-2}$ are a factor of a few larger
than those of [Ne~III] and [Ne~I], and the total infrared Ne ejecta (from [Ne~I] to [Ne~V])
are a few 10$^{-3}$ - 10$^{-2}$ M$_{\odot}$.
The uncertainty in the Ne mass is due to uncertainties in the pressure and magnetic field in
ejecta clumps.

\subsection*{4. N132D} 

\begin{figure}[!h]
\hbox{
\psfig{figure=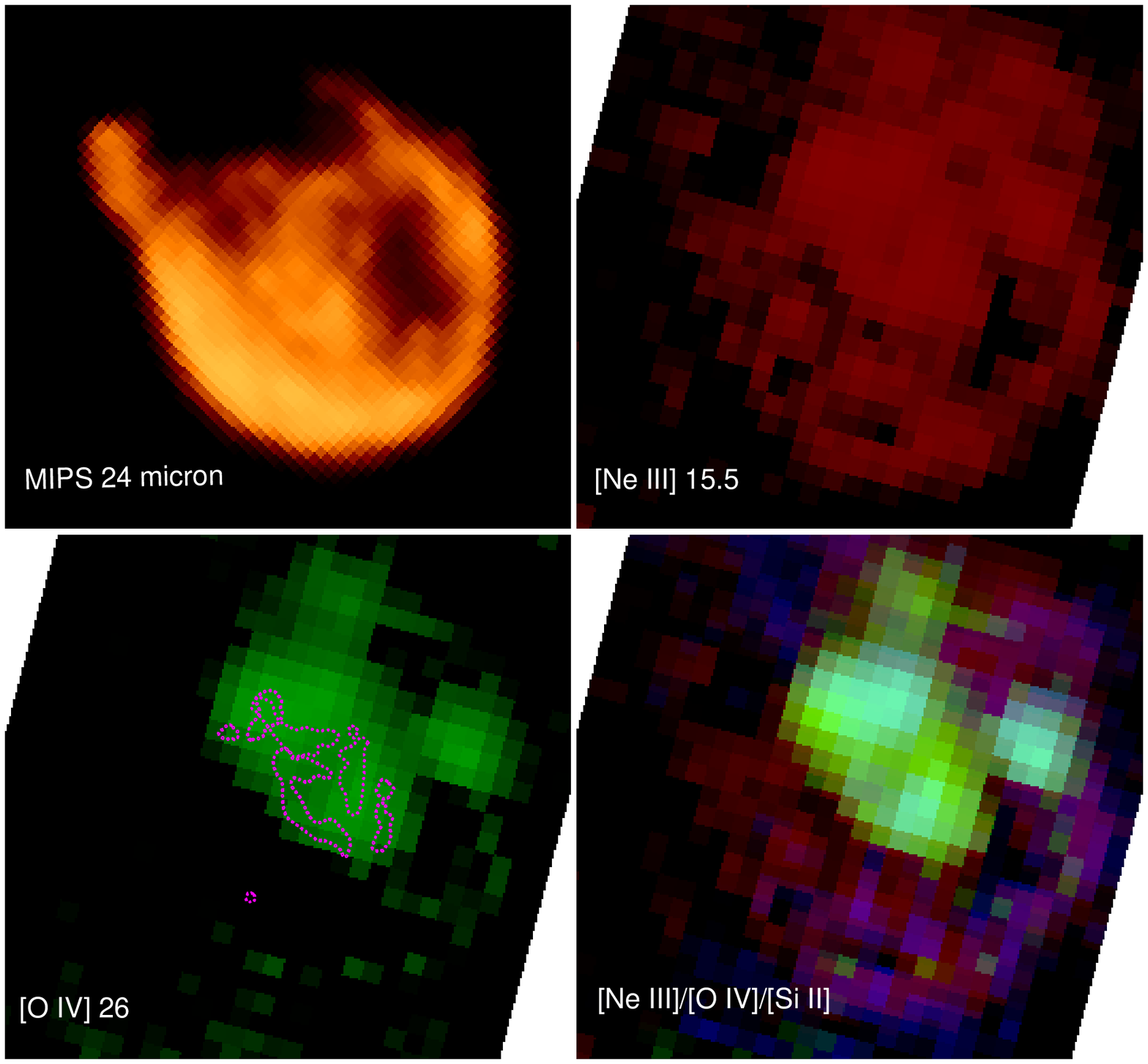,width=9truecm}
\vbox{
\psfig{figure=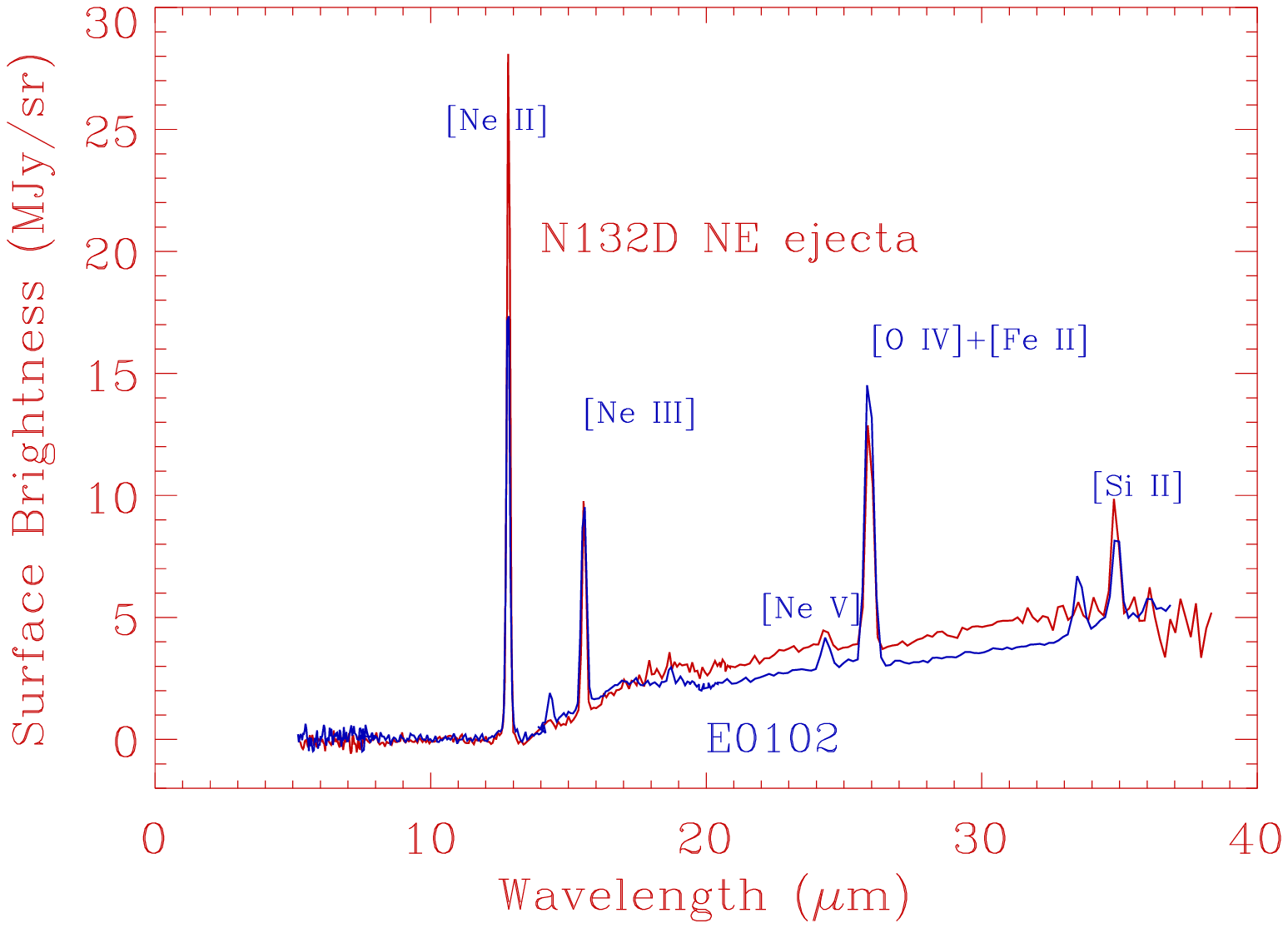,width=6truecm}
\psfig{figure=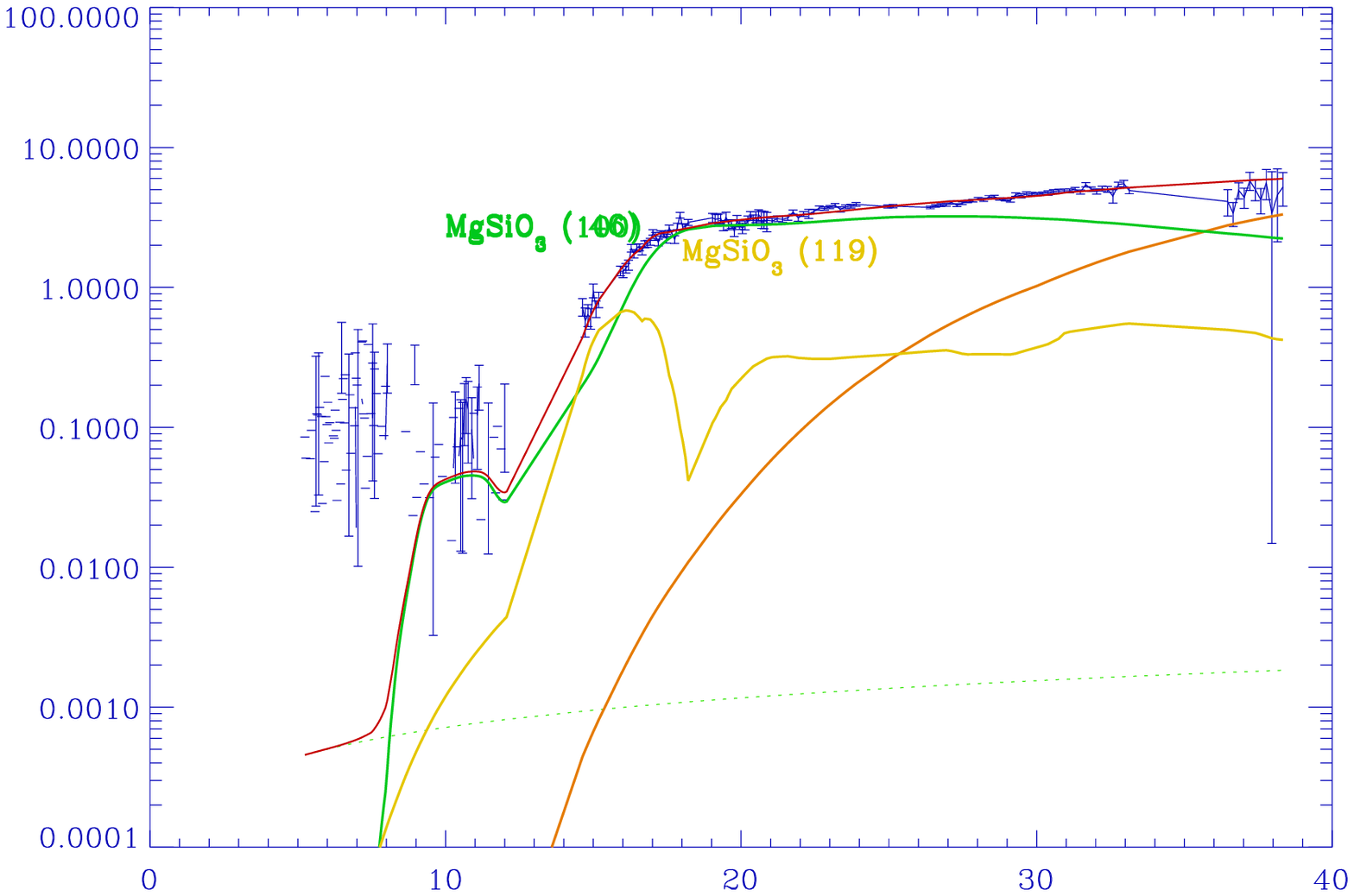,width=6truecm}
}}
\caption{(a: left) N132D {\it  Spitzer} images: MIPS 24\mic, [Ne II] (red), [O IV] (green), and 
color composite (red: [Si II]), showing that Ne, O, and Si maps are dominated by ejecta,
and coincides with the optical O-rich ejecta \citep{morse95}.
The contours show
the location of optically emitting O-rich ejecta superposed on [O IV] map.   
(b: top right) IRS spectrum of N132D (red) compared with that of E0102 (blue), showing remarkably similar
ejecta composition and dust continuum shape. (c: bottom right) The dust continuum superposed on the best fit
of dust model, a combination
of MgSiO$_3$, solid Si and carbon.} 
\label{n132dejectamap}
\end{figure}

N132D is the youngest SNR known in the Large Magellanic Clouds (LMC)
with a diameter of 100$''$ (physical size of 25 pc) and an age of $\sim$
3200 yr. N132D shows two types of emission, ejecta and ISM emission.
Optical observations identified O-rich ejecta (like Cas A) emission, an
inner ring morphology, which shows elevatered abundances and red and
blue-shifted oxygen ejecta with a velocity as high as 4400 km s$^{-1}$
\citep{morse95}.  Ejecta are oxygen-rich like in Cas A. The progenitor
is  suggested to be Type Ib,  a Wolf-Rayet (W/O) star with a mass
between 30 and 35 M$_{\odot}$ \citep{blair00}.   Recently, X-ray
observations identified highly ionized oxygen ejecta
\citep{borkowski07}. The bright shell is dominated by shocked ISM,
particulary bright in the southeastern shell \citep{blair00}. Our
previous {\it Spitzer} observations  revealed  PAH emission   from the
dominant component of 15-20 PAH hump which is interpreted as PAHs with a
relatively large number of carbon atoms \citep{tappe06}. 

We made follow-up observations (PI: Tappe, PID: 30372) with Spitzer for 
IRS-LL mapping covering the entire SNR, and IRS staring of SL
observations toward 4  positions. Staring mode was 12 cycles of 60 sec
exposure for SL. IRS-LL mapping is 2 cycle of 30 s exposure for 10$''$
stepsize, resulting in 4 minutes integration per position.  The
observations took place on 2007 Jun 23.

We present detection of infrared ejecta in N132D as shown in Fig.
\ref{n132dejectamap}a. The line maps of [Ne III] at 15.5$\mu$m and [O
IV] 26 $\mu$m, and [Si II] 34.8$\mu$m maps show strong emission at the
central ring,  which coincides with the known optical and X-ray ejecta.
Fig. \ref{n132dejectamap}b shows comparison between N132D and E0102
ejecta spectra; showing remarkably similar lines of Ne and O and the
continuum shapes. The lines at the ejecta positions are
a few ten times stronger than those in the shocked ISM position, and   
N132D shows a higher ratio of [Ne II]/[Ne III] than that of E0102,
indicating that Ne is less ionized.  We performed
spectral fitting to the IRS dust continuum as for E0102 (see
Fig. \ref{E0102irs}). The spectrum in Fig. \ref{n132dejectamap}c was fit best with the composition of
MgSiO$_3$ and solid Si and C, and the derived dust mass for the ejecta
emitting within one IRS slit is 8($\pm$3)$\times$10$^{-3}$ M$_{\odot}$.
We can also fit the spectrum with a combination of MgSiO$_3$, Si, carbon and/or
Al$_2$O3, but the mass could not be well constrained. 
This will be a lower limit of freshly formed mass, since the rest of the ejecta
and fresh dust has been mixed with shocked ISM material.

\subsection*{5. G11.2-0.3} 

G11.2-0.3 is the remnant of the historic supernova of AD 386 \citep{reynolds94}. An X-ray emitting 65 ms pulsar is located
at the center of the remnant \citep{kaspi01} with an age of 1600 yr.  
The pulsar wind nebula shows elongated, hard X-ray structures.
The distance is 4.4 kpc, based on H I absorption \citep{radhakrishnan72}.
Radio observations show bright radio emission, a circular shell of mean diameter of 4.4$'$,
22 Jy at 1 GHz with an spectral index of 0.6 \citep{green88}.

We observed IRS spectra in staring mode toward two positions  
($18^{\rm h} 11^{\rm m}
34.76^{\rm s}$, Dec.\ $-19^\circ$26$^{\prime} 30.0^{\prime \prime}$;
$18^{\rm h} 11^{\rm m}
32.32^{\rm s}$, $-19^\circ$27$^{\prime} 10.5^{\prime \prime}$).
The observations took place on 2007 October 4,
and the Long Low (LL: 15-40 $\mu$m) IRS data were taken 
with 3 cycles of 30 sec exposure time; this yields a
total exposure time of 180 sec for the first and second staring
positions.  The Short Low (SL: 5-15 $\mu$m) IRS observations were made
with 8 cycles of 14 sec exposure time; this yields a total exposure
time of 224 sec per sky position.
Reference positions were taken with IRS mapping
toward nearby positions of $\sim$30$'$ away. 
We also conducted near-infrared observations using Palomar telescope using 
narrow-filters of Fe and H$_2$ images.
The observations were conducted on 2007 October 22 and October 18.  

\begin{figure}[!h]
\hbox{
\psfig{figure=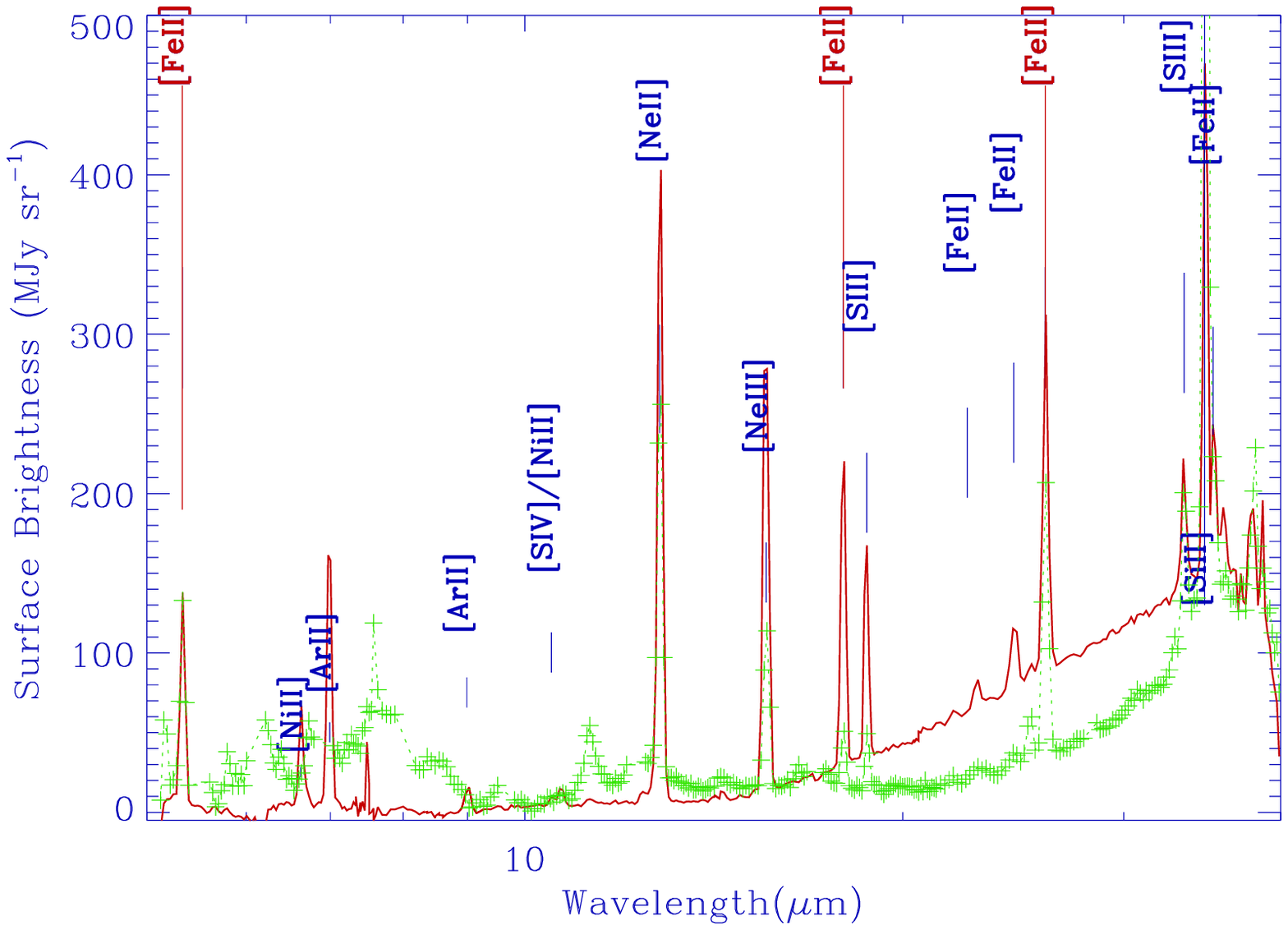,height=5.1truecm}
\psfig{figure=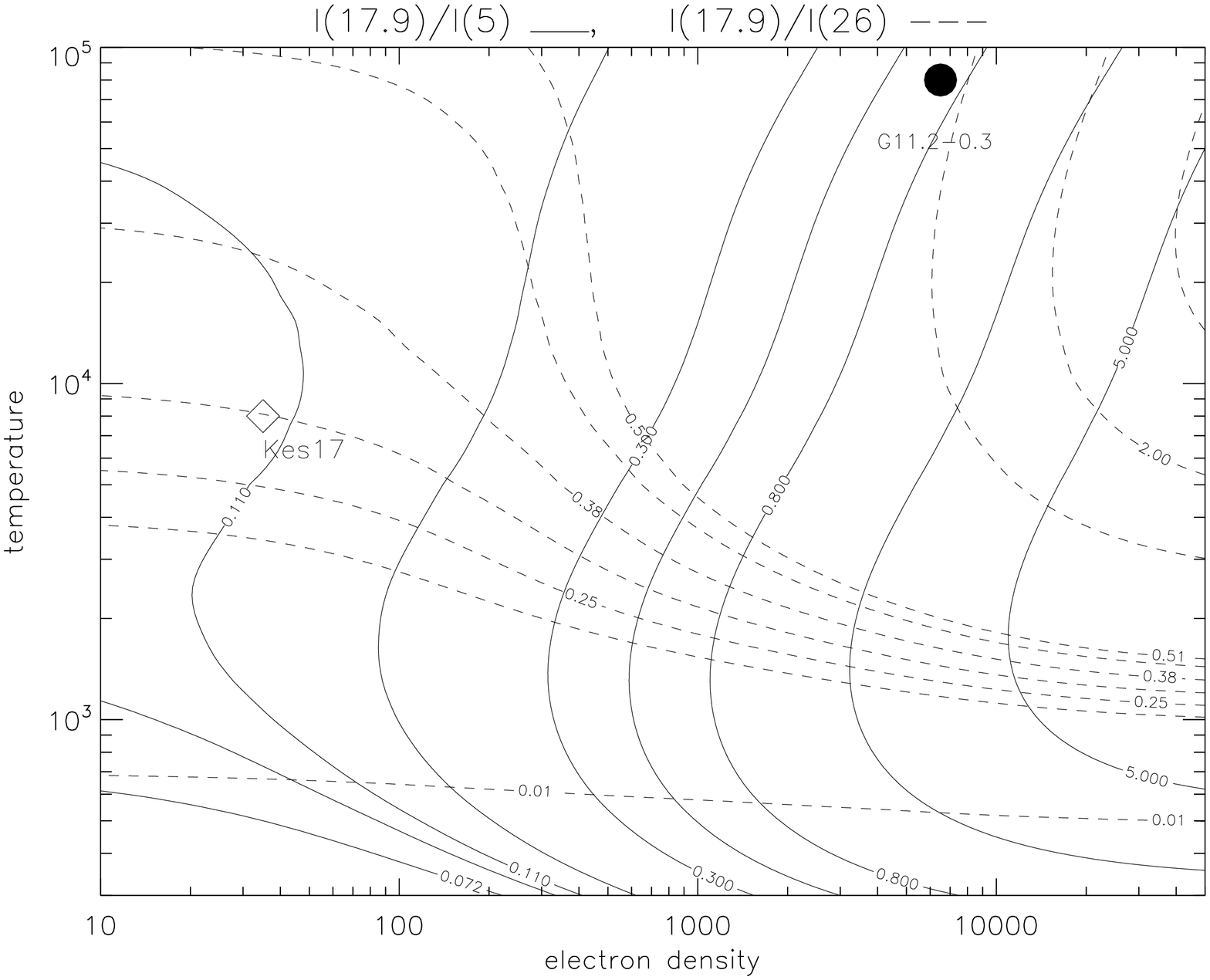,height=5.5truecm,angle=90,width=5.5truecm}
}
\caption{(a: left) The IRS spectrum of G11.2-0.3 (red) compared with that of Kes 17 (green: from Hewitt et al. 2009), 
a typical ISM dominated
SNR, showing different ratios of  5.3, 17.9 and 26 $\mu$m [Fe II] lines. 
(b: right) Contour plots of of [Fe II] diagnostic line ratios using 5.3, 
17.9 and 26 $\mu$m lines. The
observed ratios of G11.2-0.3 (filled dot) and Kes 17 (diamond) are marked, showing
that Fe emission from G11.2-0.3 
has a higher density and temperature.} 
\label{g11p2irs}
\end{figure}

Fig. \ref{g11p2irs} shows the IRS spectrum of G11.2-0.3 towards Fe-rich knots, 
showing strong \feiif\ and Ne lines, and weak Ar, 
Ni, Si and S lines. 
Diagnostics of the [Fe II] lines use excitation rate equations which are presented
in Rho et al. (2001) and we updated atomic data \citep{ramsbottom07}.
 Figure \ref{g11p2irs}b shows contours of line ratios of
17.9/5.35$\mu$m and 17.9/26$\mu$m; the former ratio is mainly sensitive to density
and the latter ratio changes depending on both density and temperature.
The two ratios imply that Fe emission of G11.2-0.3 has a temperature of $\sim$80,000 K,
and an electron density of 6000 cm$^{-3}$; such as high density is a characteristics of
ejecta. In comparison, the \feiif\ line implied electron density
from the SNR Kes 17 is  $\sim$100 cm$^{-3}$ \citep{hewitt09}. Fe is a rare element
in the circumstellar medium, but it is rich element of nucleosynthetic yields. 
The continuum of G11.2 shows smooth featureless shape 
indicates possible Fe dust.
Detection of Ni lines also implies that they are ejecta from nucleosynthetic yields.

\subsection*{6. Carbon Monoxide in the Cassiopeia A Supernova Remnant}
 
 We report the likely detection of  near-infrared 2.29 $\mu$m first
overtone Carbon Monoxide (CO) emission from the young supernova remnant
Cassiopeia A (Cas A).   The continuum-subtracted CO filter map reveals
CO knots within the ejecta-rich  reverse shock.  We compare the first
overtone CO emission with that found in the well-studied supernova, SN
1987A and find $\sim$30 times less CO in Cas A.  The presence of CO
suggests that molecule mixing is small in the SN ejecta and that
astrochemical processes and molecule formation may continue at least
$\sim 300$ years after the initial explosion. This section is a summary
of a Journal paper \citep{rho08casaco}.

\noindent{\bf Introduction and Observations}
Supernovae (SNe) are suggested to be the first molecular factories in 
the early Universe \citep{cherchneff08}.   Understanding the chemistry
within the ejecta, and the abundance and mixing of CO and SiO
molecules, is particularly important in studying dust formation
processes since this controls how much carbon is available to form
amorphous C grains and how much  oxygen available beyond that bound in
CO and SiO molecules to form MgSiO$_3$ and  Mg$_2$SiO$_4$.

\begin{figure} 
\hbox{
\psfig{figure=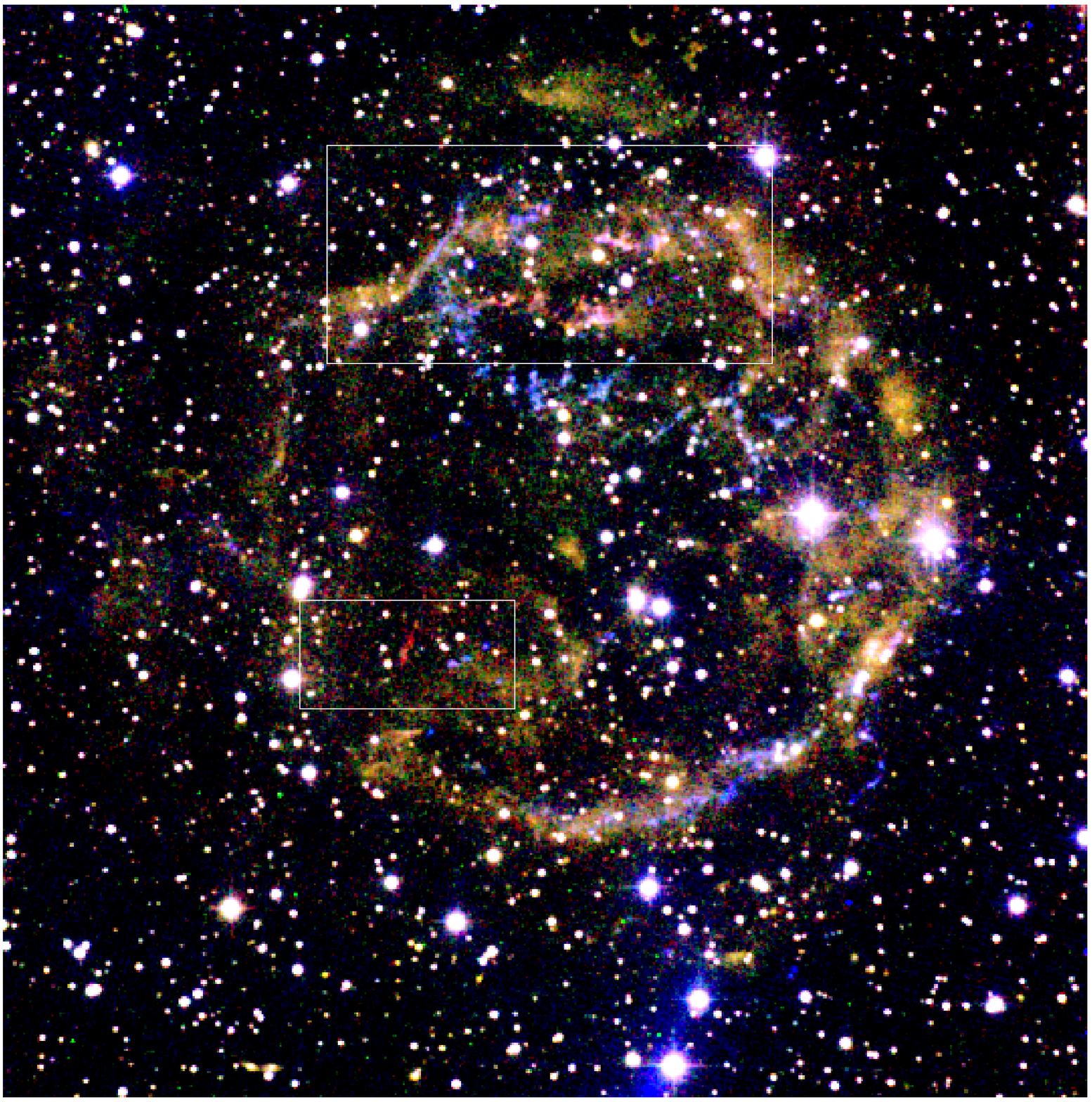,width=5.5truecm}
\psfig{figure=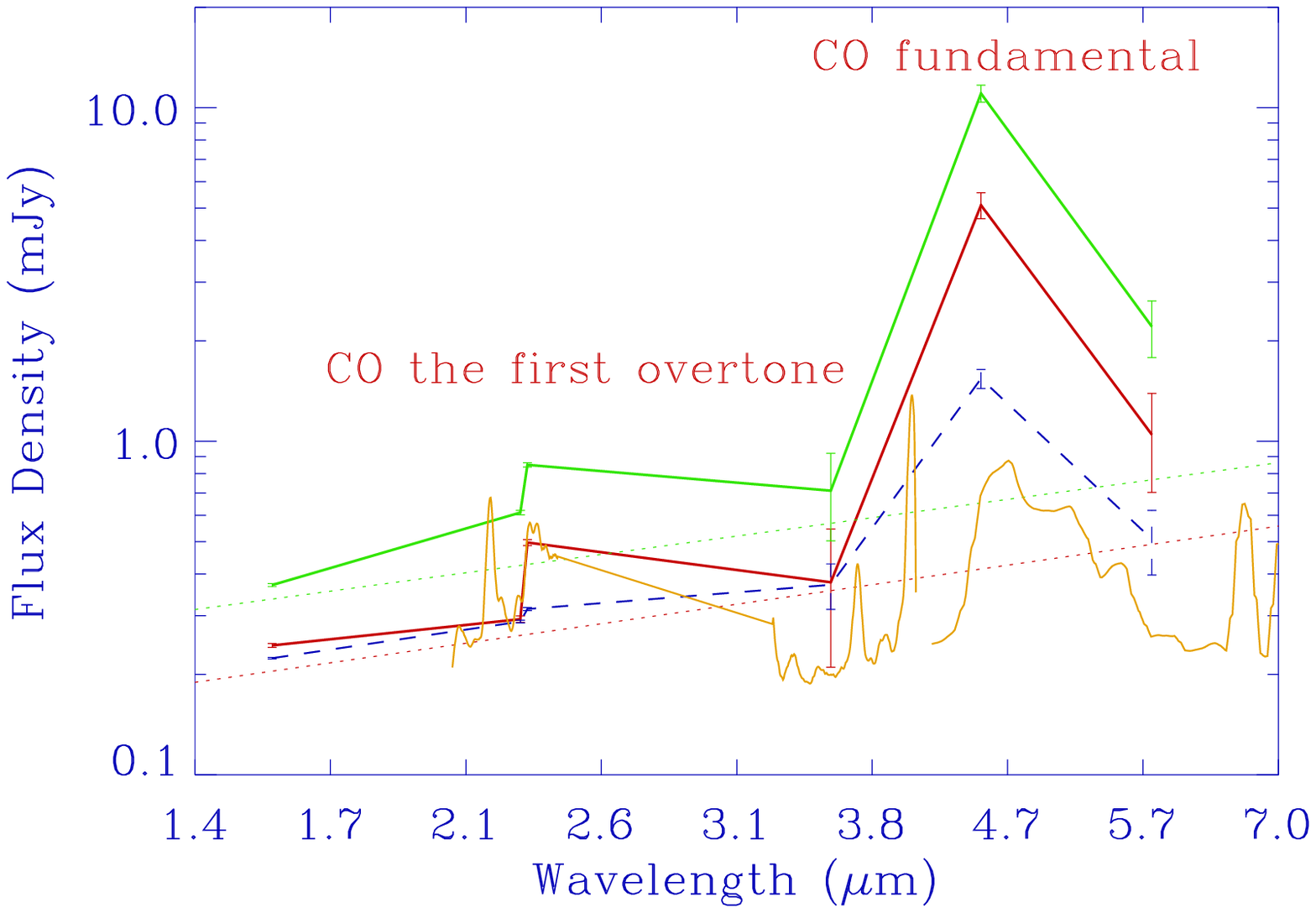,width=6truecm}
}
\caption{(a: left) Mosaicked color image of Cas A with WIRC, with CO (red),
K-continuum (green) and P$\beta$ (blue).   The range of surface
brightnesses measured from the diffuse structures are 0.31 - 2.79$\times$10$^{-5}$ for CO,  0.23 - 1.4$\times10^{-5}$ for K-continuum, and 1.2 - 29.0$\times10^{-5}$ erg
s$^{-1}$ cm$^{-2}$ sr$^{-1}$  for P$\beta$. CO-excess regions (in
red) are marked as boxes (the bottom box marks the location of the Minkowski knot). 
(b: right) The spectral energy distribution of two representative CO
emitting knots: knot1 (green) and knot2 (red) 
which show excess emission at 2.294 \mic.   The
contribution from the synchrotron continuum is shown by the dotted
line. Comparison with the SED (scaled to the CO knot1 continuum) of the
NW ejecta knot (blue dashed curve) is shown.
The SN 1987 spectrum is shown for comparison (orange curve; Meikle et al. 1989).}
\label{casacothreecolor} 
\end{figure}

We observed Cas A on 2006, September 3 and 4 with the Wide field
InfraRed Camera (WIRC) on the Hale 200 inch (5 m)
telescope at Mount Palomar.  We took narrow-band images using filters
centered on CO(2-0) at 2.294 \mic, K-band and H-band continuum at 2.270
and 1.570 \mic, and P$\beta$ at 1.182 \mic\ (see Figure
\ref{casacothreecolor}).  The exposure time of CO, K-cont and P$\beta$
is a has a total on-source integration time of 8100 sec for CO, 7290
sec for K-cont and 810 sec for P$\beta$ image. 

\noindent{\bf Results and Discussion} 
Fig. \ref{casacothreecolor}a shows a color composite image of Cas A,
combining the P$\beta$ line (blue), 2.27 \mic\ continuum (green), and CO
(2-0) 2.29 \mic\ line (red).    The 2.27 \mic\ continuum consists mostly
of non-thermal synchrotron radiation and delineates the shell structure
(see Rho et al.  2003). Finally, the CO (2-0) overtone emission traces
the dense knots of gas along the shell and interior interface. The
molecular emission is evident as red in Figure \ref{casacothreecolor}
and more easily seen in the continuum-subtracted CO image, revealing the
line emission towards the northern parts of bright rings  and an eastern
filament of the Minkowski knots in the southeast. The CO emission is
highly clumpy and distributed in a collection of knots.  The locations
where CO emission is detected coincide with the ejecta at the reverse
shock   \citep{hwang04,rho08}. It indicates that the CO
gas was formed in the ejecta.  
We compiled spectral energy distributions (SEDs) of near-infrared H,
K-continuum, CO bands and \spitzer\ IRAC Bands 1, 2 and 3 (3.6 - 5.8 \mic)
for bright
knots;  Fig. \ref{casacothreecolor} shows three representative SEDs after
correcting for the extinction. The SEDs show a sudden jump at 2.294
\mic\ toward the CO detection positions, with no such jump toward the
NW ejecta position.  The SEDs of the CO knots also show a larger IRAC
Band  2 (4.5 \mic) excess than that of the NW ejecta knot.

We estimated the mass of CO in Cas A by comparing the CO flux density in
Cas A with that found in the supernova SN 1987A. Fig.
\ref{casacothreecolor} compares the CO emission from Cas A and  SN 1987A
\citep{meikle89}. After accounting for the distance differences,  the CO
(2-0) first overtone flux density of Cas A is a factor of $\sim 28$
smaller  than that of SN1987A on day 255.   Assuming the same gaseous
conditions as seen in the CO emitting regions in SN1987A (a temperature
of 1800-2800 K and a velocity of 2000 km~s$^{-1}$), we estimated the CO
mass in Cas A to be  $\rm M_{CO}\sim (1.7-8.2)\times 10^{-6} \,\rm
M_{\odot}$. The CO detections in SNe have been within a few years after
the initial explosion.  Our CO detection suggests that either the  CO
layer is not macroscopically mixed with ionised helium or that the 
helium is not ionised.   Typical electron density of C- and O-rich
regions during formation of CO is shown to be 5$\times$10$^8$ cm$^{-3}$
\citep{gearhart99}.   \cite{travaglio99} assumed that mixing occurs   at
the molecular level, prior to dust condensation, while Clayton \&
Nittler (2004), Deneault, Clayton \& Heger (2003) suggest that gases are
unlikely to be mixed at the molecular level within a few years. Our CO
detection suggests that molecular mixing is small (at least smaller than
previously thought) in the SN ejecta and during the development of the
reverse shock.


\noindent{\bf 7. Conclusion} We  present unambiguous  evidence of dust
formation in young SNRs. The dust map is almost  identifical to the ejecta
map, showing that dust is formed  in the ejecta.  Identified dust
composition in SNRs  include the features of SiO$_2$ [21 and 9 $\mu$m],
MgSiO$_3$ [17 and 9.5 $\mu$m], Si [15 $\mu$m], SiC [11.3 $\mu$m] and the
featureless dust of carbon, Al$_2$O$_3$, and Fe. We also show that dust
and ejecta compositions are closely correlated. E0102 shows dominant
ejecta of Ne and O, while Cas A shows Ar, Si and Fe ejecta in additional
to Ne and O. A total mass of freshly formed dust is
0.02- 0.054 M$_{\odot}$ for Cas A, and 0.015 M$_{\odot}$  for E0102.
G11.2 also shows similar order of dust mass including Fe dust. The dust mass
is more than one order higher than previous estimates but slightly lower
than theoretical predictions.  CO molecule detection from Cas A indicates
on-going processes of astro-chemisty. 

\acknowledgements 
 This work is based on observations made with the \spitzer\ {\it
Space Telescope}, which is operated by the Jet Propulsion Laboratory,
California  Institute of Technology.
Support for this work was provided by NASA and through LTSA grant  NRA-01-01-LTSA-013
and \spitzer\ GO awards issued
by JPL/Caltech.



{}


\end{document}